\documentclass[conference]{IEEEtran}
\IEEEoverridecommandlockouts
\usepackage{cite}
\usepackage{amsmath,amssymb,amsfonts}
\usepackage{algorithmic}
\usepackage{graphicx}
\usepackage{textcomp}
\usepackage{xcolor}
\usepackage{enumitem}
\def\BibTeX{{\rm B\kern-.05em{\sc i\kern-.025em b}\kern-.08em
    T\kern-.1667em\lower.7ex\hbox{E}\kern-.125emX}}
\begin{document}

\title{Study of software developers' experience using the Github Copilot Tool in the software development process.\\

\thanks{Identify applicable funding agency here. If none, delete this.}
}

\author{\IEEEauthorblockN{1\textsuperscript{st} Mateusz Jaworski}
\IEEEauthorblockA{\textit{Warsaw University of Technology} \\
Warsaw, Poland}
\and
\IEEEauthorblockN{2\textsuperscript{nd} Dariusz Piotrkowski}
\IEEEauthorblockA{\textit{Warsaw University of Technology} \\
Warsaw, Poland}
}

\maketitle

\begin{abstract}

In software development there is a constant pressure to produce code faster and  faster without compromising on quality. New tools supporting developers are created in response to this demand. Currently a new generation of such solutions is about to be launched - Artificial Intelligence driven tools. On 29 June 2021 Github Copilot was announced. It uses trained model to generate code based on human understandable language. The focus of this research was to investigate software developers' approach to this tool. For this purpose a survey containing 18 questions was prepared and shared with programmers. A total of 42 answers were gathered. The results of the research indicate that developers' opinions are divided. Most of them met Github Copilot before attending the survey. The attitude to the tool was mostly positive but not many participants were willing to use it. Concerns are caused by security issues associated with using of Github Copilot.

\end{abstract}


\section{Introduction}
Nowadays software developers have at their disposal tools providing automatic code completion and syntax suggestions. Though these tools are helpful and greatly accelerate developers' work there is a constant pressure to produce code faster and faster. For this reason a new generation of tools is under development - machine learning-based code generators. They are trained models which generate code basing on some definition, e.g. natural language description [1,2]. 

Announced on 29 June 2021 Github Copilot is an artificial intelligence tool developed as an extension for the Visual Studio Code and JetBrains IntelliJ IDEs [3]. Github Copilot base on language model Codex, trained on large amount of source code. It suggests the source code of function basing on e.g. name of this function and given variables [1]. The programmer is given with some recommended implementations of required part of code and they are able to choose the most suitable one and adjust it if necessary.

In this work we make a research of software developers' experience with using Github Copilot and their approach to this tool and its future. 

\subsection{Problem Statement}

Currently we are witnessing transformation of developers supporting tools. The state of the art AI algorithms allow not only to suggest or auto-complete syntax but also to generate entire parts of code. Till now there were no comparable tools which would help developers so much. One of the first so advanced software is released on 29 June 2021 the new AI-based code generator, Github Copilot. It is much more advanced than widely used tools providing code completion and syntax suggestions. The development of such tools will probably influence on both developers daily work and their situation on labor market. 

\subsection{Objective}

This work aims to build an understanding of developers' experience and attitude to Github Copilot. We are trying to learn how they see usefulness of this tool in their daily work, what encourage them to use it and how they see the influence of Github Copilot on employment in IT.

\subsection{Contribution}

For the purpose of this study a total of 42 answers from developers with different seniority level and specialization were examined. The survey consisted of questions associated with general attitude of developers to Github Copilot, their prediction about future of such tools and their impact on software engineering and security of this solution.

\section{Related Work}

As Github Copilot is a new solution released less than one year ago there was not much research conducted on it. However scientists have been studying for years the use of Artificial Intelligence in programming. One of the method - genetic programming - was compared to Github Copilot by Dominik Sobania, Martin Briesch and Franz Rothlau in their 'Choose your programming Copilot' [1] work. They studied the performance of both solutions applying Github Copilot to the common program synthesis benchmark problems. The achieved results were then compared with results of genetic programming algorithms described in related literature. The research showed similar performance of both tools in solving popular algorithmic problems.

The use of artificial intelligence in programming is also the topic of Raphael's Jenni work 'Machine learning for Programming Languages'[4]. The author gives a general overview of application of deep learning in supporting developers in coding. As one of the modern tools using this technology Raphael Jenni mentions Github Copilot.

Stuart Fitzpatrick in his work 'On the Nature of AI Code Copilots' also describes some aspects of Github Copilot. He considers if AI pair programmers constitute a ‘compiled’ form of the training data, similar to an executable file, or if they are closer to the source code of a conventional program and can be modeified by further traingning [5]. Fitzpatrick brings up also the matter of copyrights of the code used by Github Copilot in its process of searching suggested solutions to described problem.

Neil A. Ernst and Gabriele Bavota describe AI driven development environments as a new revolution in software engineering in their work 'AI-driven Development Is Here: Should You Worry?'[3]. They distinguishes three eras in programming: the first one when programmers had to write entire code on their own, the second one when IDEs suggest syntax and provide autocompletion and the third one - which is underway - when AI-driven Development Environments generate code. As an example of such environment the authors give Github Copilot. The work includes short introduction to the tool, describes what are expectations from this kind of environments and what challenges are facing them.

When talking about Github Copilot, security issues of generated code causes a lot of concernes. The tool uses model trained on open source Github repositories. The code included in these repositories is often unvetted so it is possible that the model was learned on buggy code. This is a subject of study conducted by five authors of 'Asleep at the Keyboard? Assessing the Security of GitHub Copilot’s Code Contributions' work [2]. They generated with Github Copilot solutions for security sensitive problems and evaluated the results. Out of 1689 generated programs about 40\% were recognized as vulnerable.

\section{Research Design and Methodology}

For the purpose of this study a survey was designed and prepared. All questions were created by authors who are software developers with experience in programming. Then the questions were reviewed and the remarks were taken into account in the final version. The survey included 18 questions, 14 out of them were closed-ended and 4 open-ended. The survey was divided into 3 sections. First section included 2 questions asking if a person heard about Github Copilot or used it. Next 11 questions investigated an attitude of participant to the tool. This section characterized also a participant by asking about their experience, specialization and used technology stack. The third part of the survey refers to the topic of security of Github Copilot.

The survey was created in Google Forms and distributed between software developers from the authors' environment and through social media including Facebook and Linkedin. All closed-ended questions were obligatory and open-ended questions were optional. No fake answers were detected so all of them were taken into consideration. A total of 42 answers were gathered in a period from 07.03.2022 to 25.04.2022. The results were exported to Excel and figures were created using available tools.

\section{Results}
This section presents 42 results of the survey. The first subsection provides general overview whether programmers have heard about the tool and their attitude towards it. The next part presents data on the types of programmers and their experience. The last subsection deals with the topic of security and changing the attitude of the respondents after reading the information from the creators of GitHub Copilot about the security of the tool.

\subsection{General overview}
\begin{figure}[htbp]
\centerline{\includegraphics[scale=0.35]{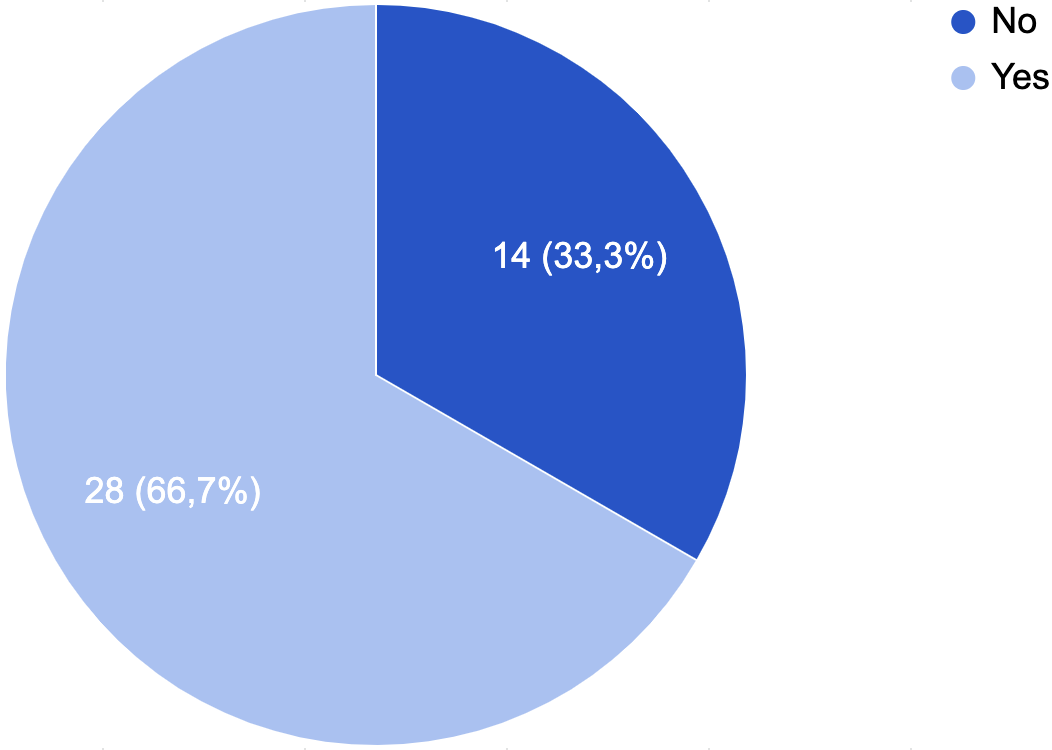}}
\caption{Hearing about GitHub Copilot}
\label{fig:Q1}
\end{figure}

The aim of the first question is to gather information whether the developers have heard about GitHub Copilot. As shown in Fig.~\ref{fig:Q1} 28 (66,7\%) have heard about it and 14 (33,4\%) did not.

\begin{figure}[htbp]
\centerline{\includegraphics[scale=0.35]{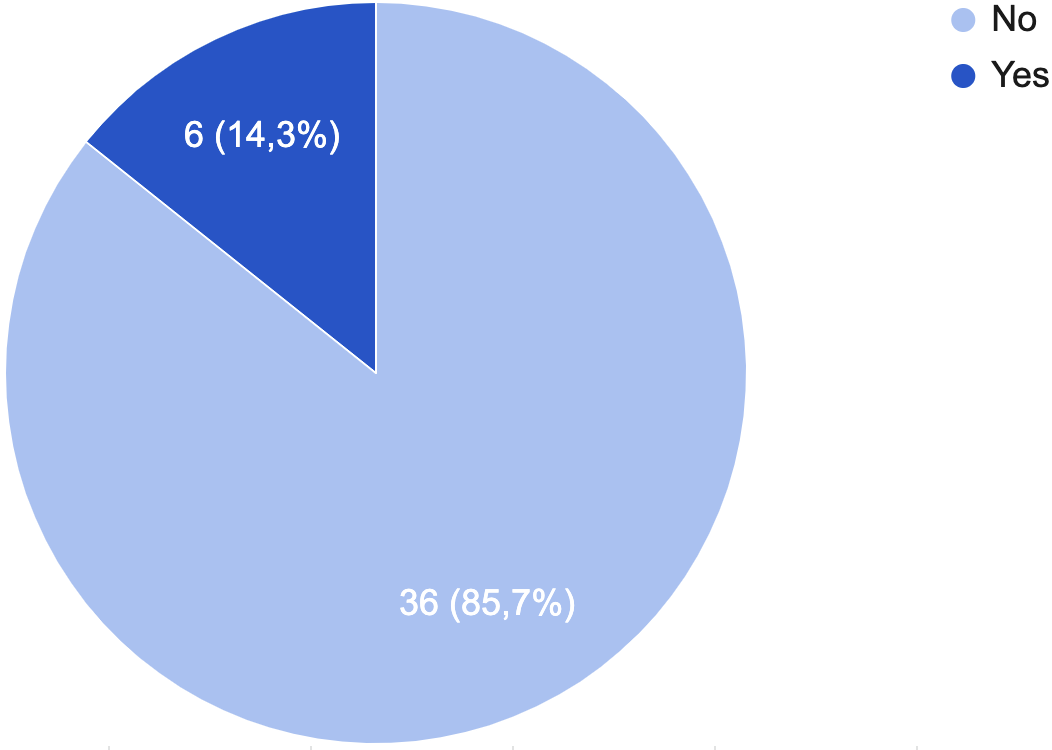}}
\caption{Usage of GitHub Copilot among the respondents}
\label{fig:Q2}
\end{figure}

What is more interesting only 6 (14,3\%) people were using it and 36 (85,7\%) were not. The result of second is shown in  Fig.~\ref{fig:Q2}

The next three questions (third, fourth and fifth) were directed to the 6 people who were using GitHub Copilot. 

The third question asked about incentive to use. They were answered: 

\begin{itemize}
    \item Speed of writing code
    \item Work
    \item Better performance
    \item Innovative solution
    \item Curiosity
    \item Facilitating the writing of repetitive code fragments
\end{itemize}

Next, in the fourth question, they were asked how long they had been using the tool and 3 (50\%) of them were using it for two weeks, the other half more than two weeks. In addition, one respondents replied: 'several months'. 

\begin{figure}[htbp]
\centerline{\includegraphics[scale=0.4]{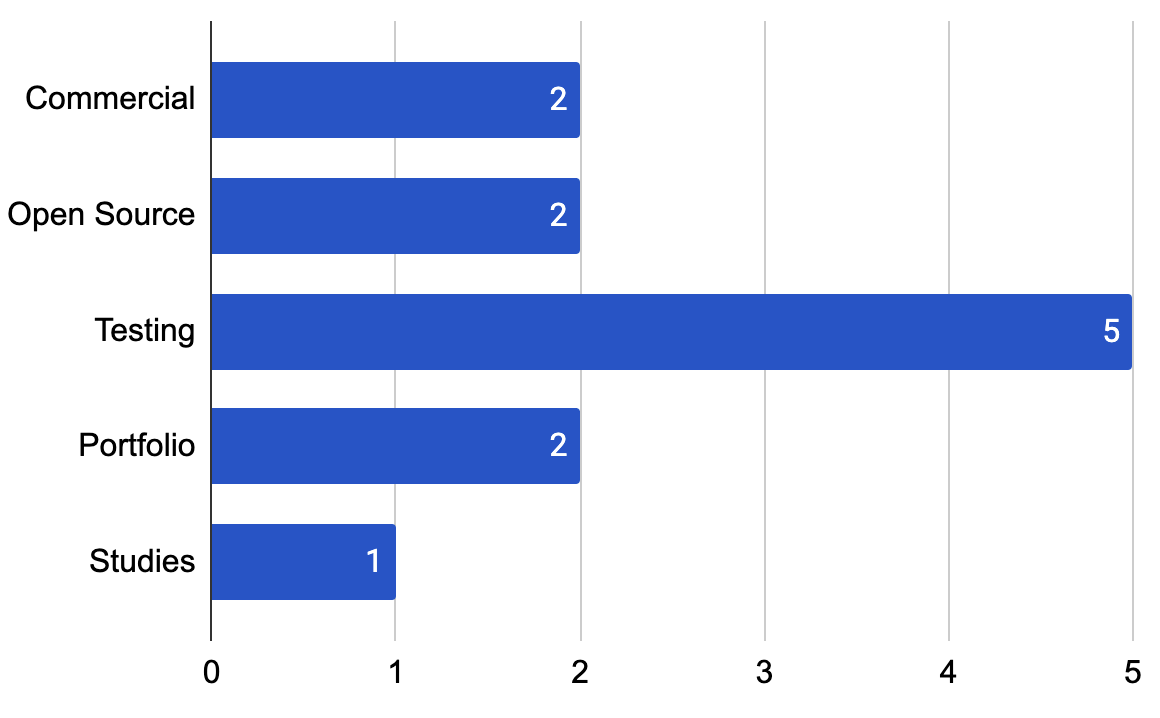}}
\caption{Types of projects in which the tool was used}
\label{fig:Q5}
\end{figure}

In Fig.~\ref{fig:Q5}, where are shown results of multiple choice question number five, are types of projects in which GitHub Copilot was applied. 5 of developers were testing it, 2 of them were using it for commercial project, open source or portfolio and 1 person for studies.

All subsequent questions were directed to the full group of respondents. 

\begin{figure}[htbp]
\centerline{\includegraphics[scale=0.4]{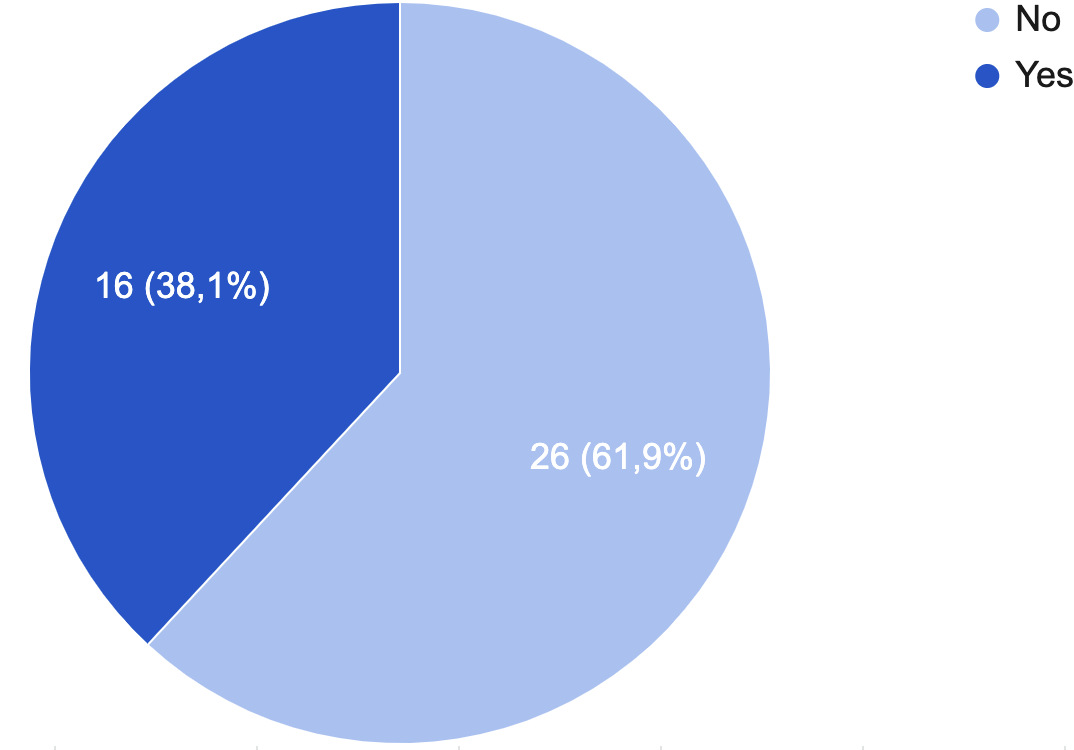}}
\caption{Intentions to test the tool}
\label{fig:Q6}
\end{figure}

In the question number six they were asked about their intentions to test the tool. 
The results are shown in Fig.~\ref{fig:Q6} and more than half, which is 26 (61,9\%) did not want to test it and 16 (38,1\%) did.

\begin{figure}[htbp]
\centerline{\includegraphics[scale=0.4]{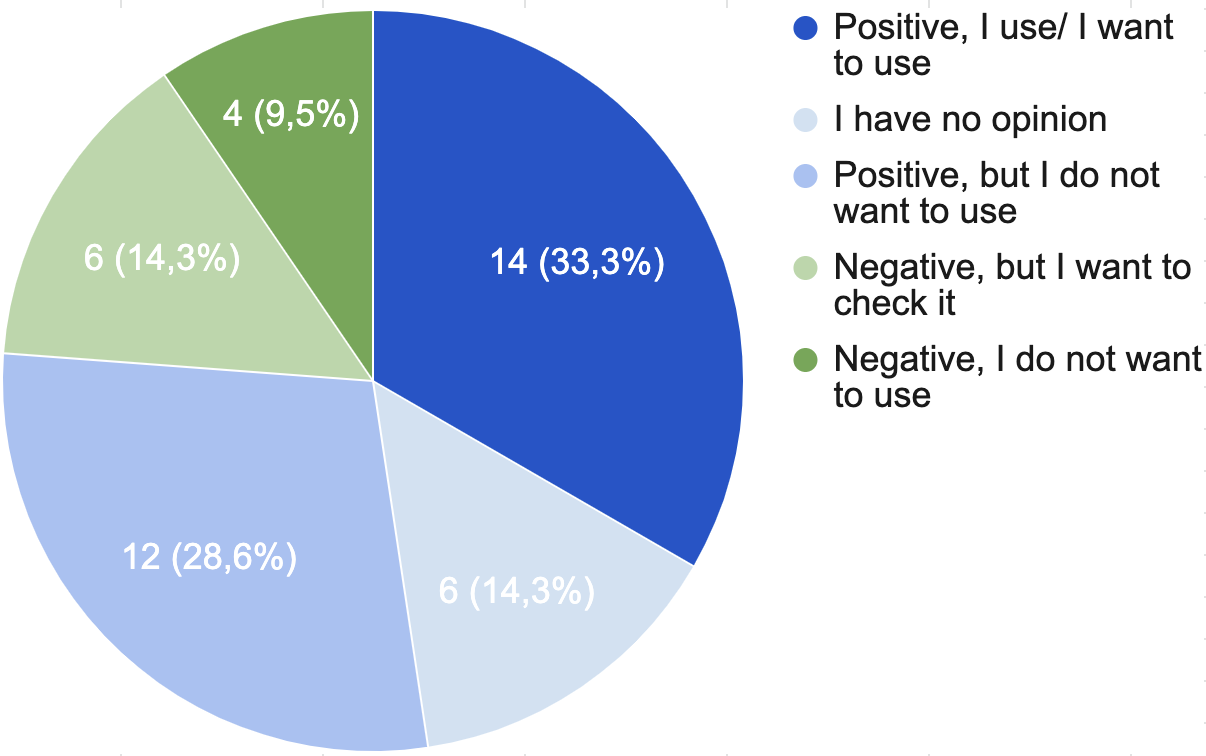}}
\caption{Attitude to that kind of tools}
\label{fig:Q7}
\end{figure}

The seventh question asked about developers' attitude to artificial intelligence tools. In Fig.~\ref{fig:Q7} their answers have been presented. 14 (33,4\%) have positive attitude and they are using it or want to use, 12 (28,6\%) they think alike but do not want to use. 6 (14,3\%) developers does not have opinion. 6 (14,3\%) of the respondents are positively minded but they do not want to use it. Only 4 (9,5\%) answers are negative and they do not want to test it.

\subsection{Participants Characteristic}
\begin{figure}[htbp]
\centerline{\includegraphics[scale=0.33]{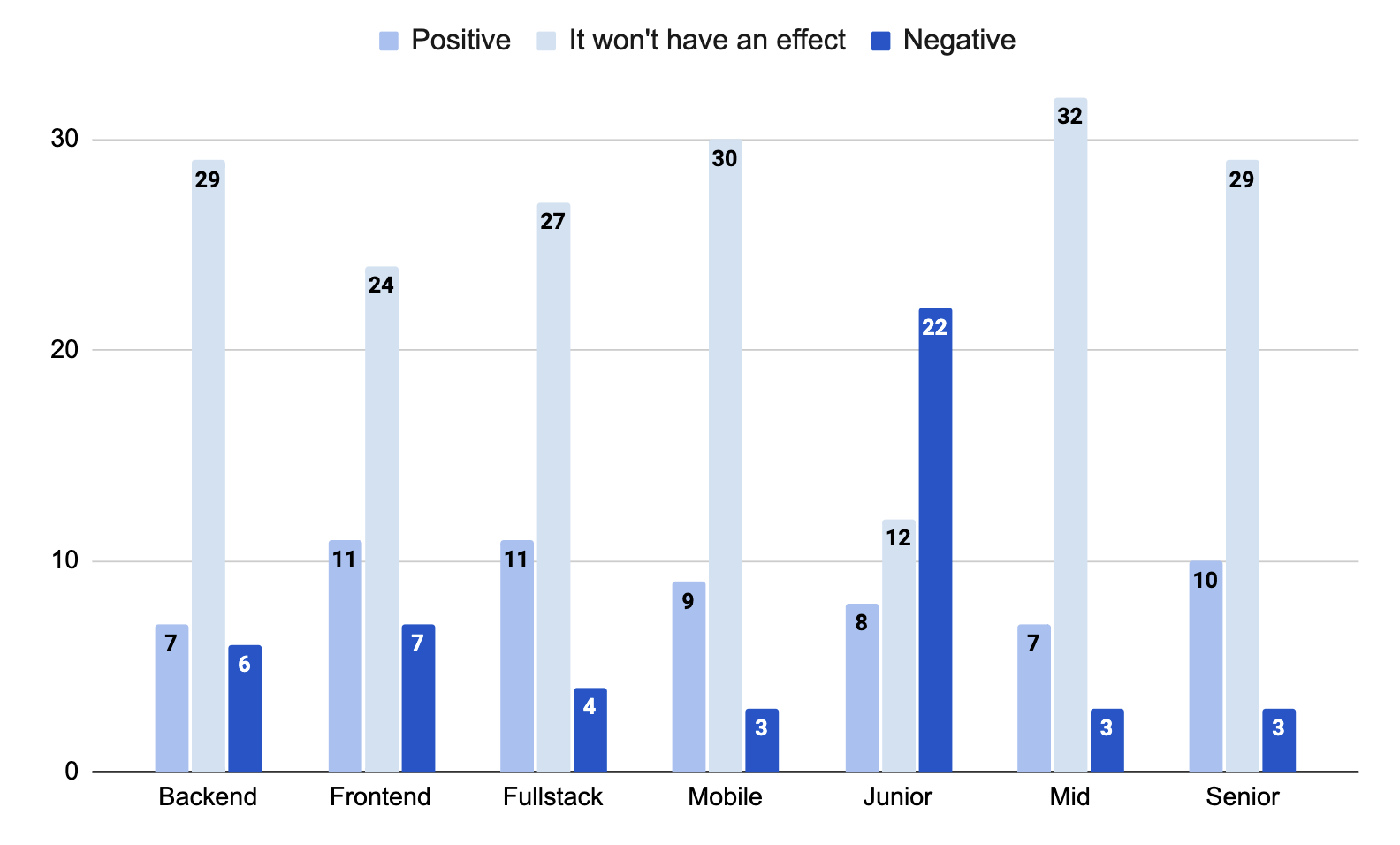}}
\caption{Impact on the future employment of programmers}
\label{fig:Q8}
\end{figure}

The second group of questions was created to check and characterize types of programmers and their experience levels. 
The eight question asked about negative impact on the future employment of programmers by different groups. In Fig.~\ref{fig:Q8} their answers have been presented. The chart can be divided into two groups: by developer type and experience level. The results are presented according to the pattern: 

{\it[Type] (Positive; It will not have an effect; Negative)}, which shows type and number of developers.
\begin{enumerate}[label=(\alph*)]
\item Experience level
    \begin{itemize}
    \item Junior (8; 12; 22)
    \item Mid (7; 32; 3)
    \item Senior (10; 29; 33)
    \end{itemize}
\item Type of developer
\begin{itemize}
    \item Backend (7; 29; 6)
    \item Frontend (11; 24; 7)
    \item Fullstack (11; 27; 4)
    \item Mobile (9; 30; 3)
    \end{itemize}
\end{enumerate}

\begin{figure}[htbp]
\centerline{\includegraphics[scale=0.41]{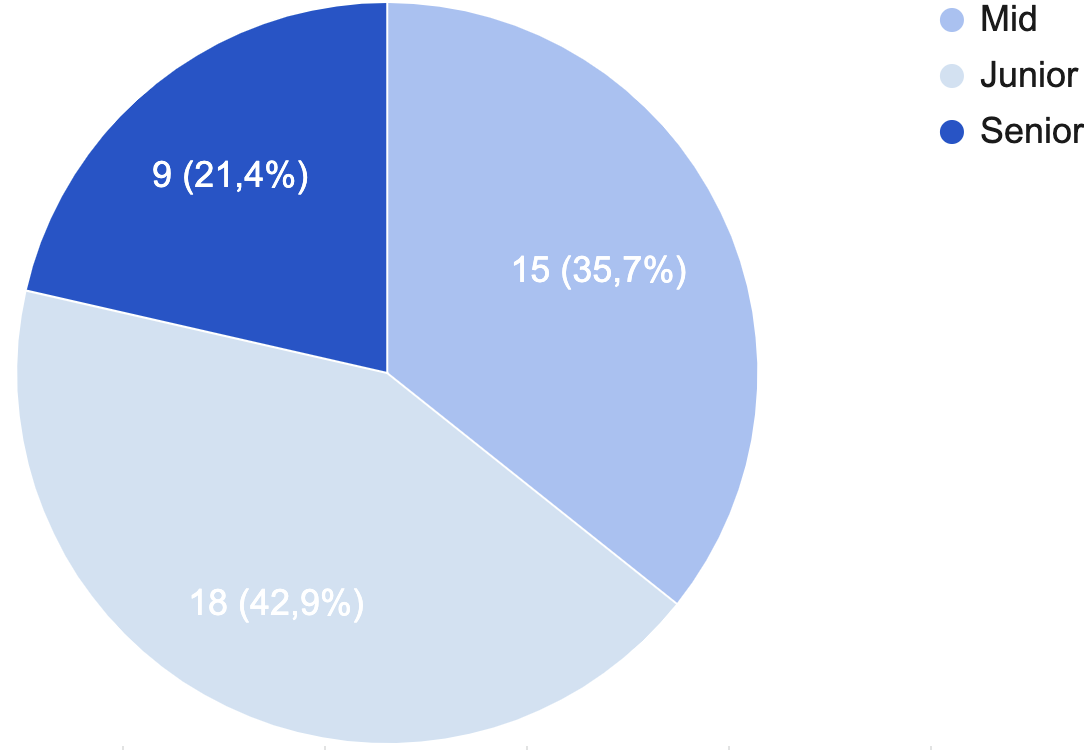}}
\caption{Experience level}
\label{fig:Q10}
\end{figure}

Next question asked about their experience level. As shown in Fig.~\ref{fig:Q10} the Juniors' responses were the most numerous 18 (42,9\%), next Mid's 15 (35,7\%) and the least Senior's 9 (21,4\%).

\begin{figure}[htbp]
\centerline{\includegraphics[scale=0.42]{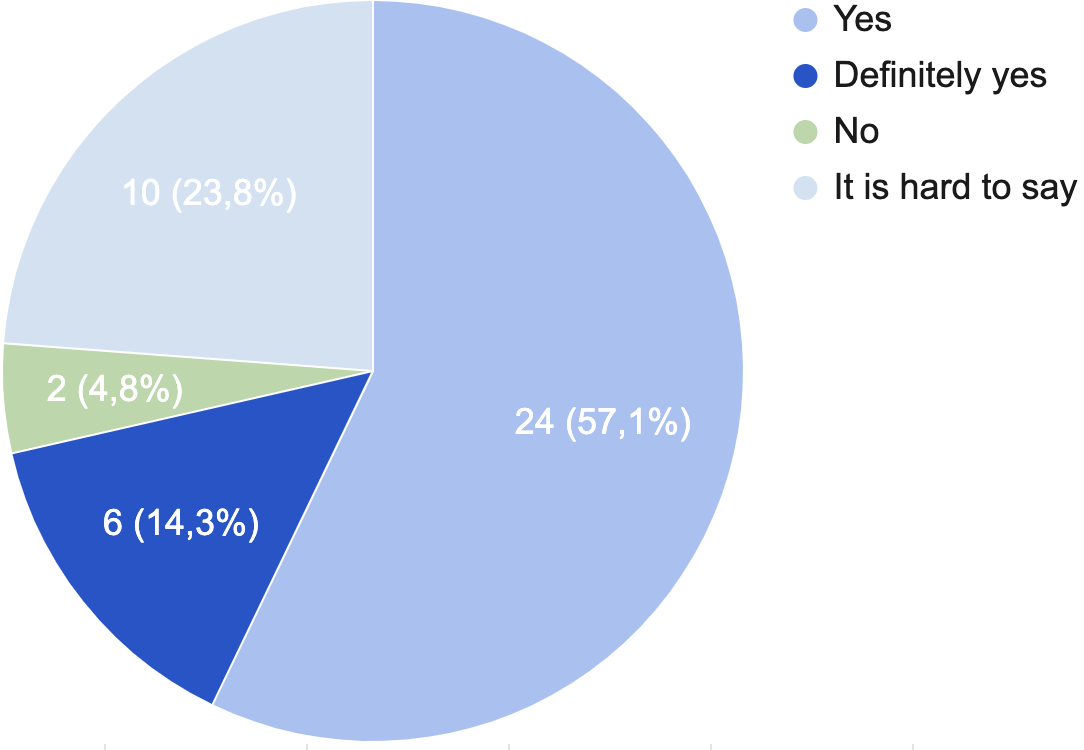}}
\caption{Helpful of GitHub Copilot}
\label{fig:Q9}
\end{figure}

The tenth question in this section participants were asked that this tool will be helpful. More than half were said 'Yes' 24 (57,1\%) and 'Definitely yes' 6 (14,3\%). For 10 (23,8\%) developers were hard to express their opinion and only 2 (4,8\%) think whether it will be helpful. The results are shown in Fig.~\ref{fig:Q9}

\begin{figure}[htbp]
\centerline{\includegraphics[scale=0.38]{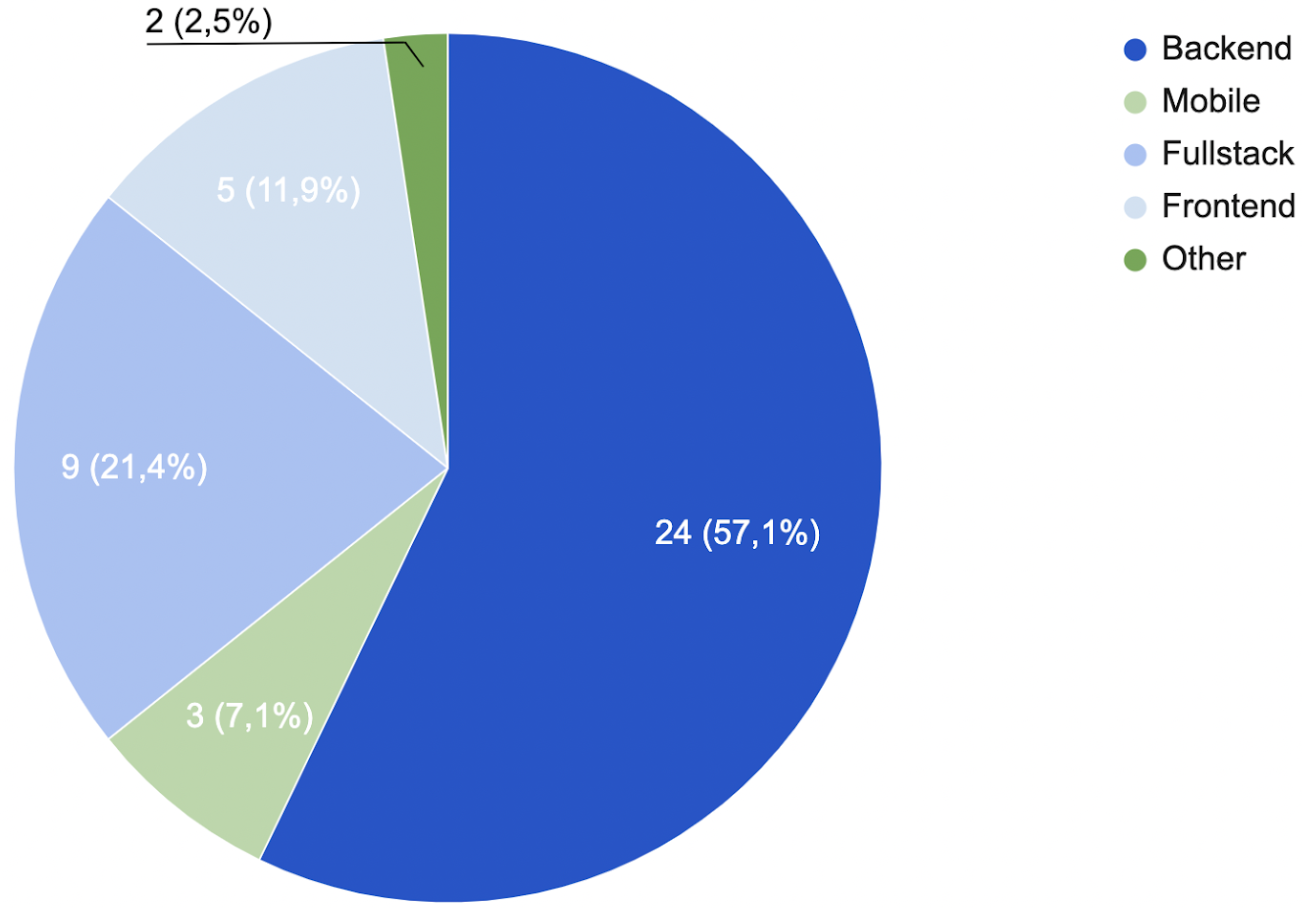}}
\caption{Developers' specializations}
\label{fig:Q11}
\end{figure}

In the eleventh question participants were asked to choose their specializations. As shown in Fig.~\ref{fig:Q11} the most are Backend developers 24 (57,1\%), then Fullstack 9 (21,4\%) and Frontend 5 (11,9\%). The least numerous group were Mobile 3 (7,1\%) and the others 2 (2,5\%). 

Next they were pleased to write their technological stack. Java received the most responses, then JavaScript and Python. There were also answers like C\# or C++, but there were much fewer of them. Developers, depending on the technology, wrote different frameworks.

\subsection{Security}

While using GitHub Copilot, code suggestions are downloaded from the internet. Firstly, respondents were asked about their concerns about the data sent ex. exploits. Then they were shown information from the authors about safety:

{\it 'All data is transmitted and stored securely. Access to the telemetry is strictly limited to individuals on a need-to-know basis.'} [5]

\begin{figure}[htbp]
\centerline{\includegraphics[scale=0.21]{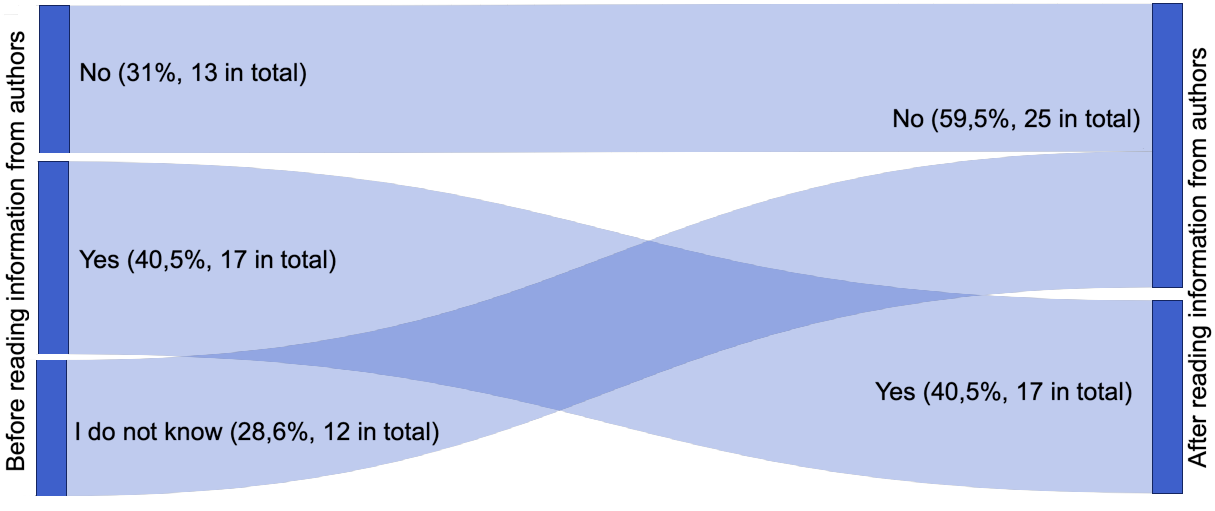}}
\caption{Concerns about data sent into project}
\label{fig:Q1415}
\end{figure}

Next they were pleased to answer again. In Fig.~\ref{fig:Q1415} are shown the results before and after reading information from the authors. Almost half of the developers' group 17 (40,5\%) are afraid of data send into project, 12 (28,6\%) do not know and 13 (31\%) trust the authors. After reading information 25 (59,5\%) people from the group have concerns and 17 (40,5\%) have not.

Secondly they were asked about concerns about data leak from private commercial repositories. And again they read information from authors then have to reply:

{\it 'We use telemetry data, including information about which suggestions users accept or reject, to improve the model. Your private code is not used as suggested code for other users of GitHub Copilot.'} [5]

\begin{figure}[htbp]
\centerline{\includegraphics[scale=0.21]{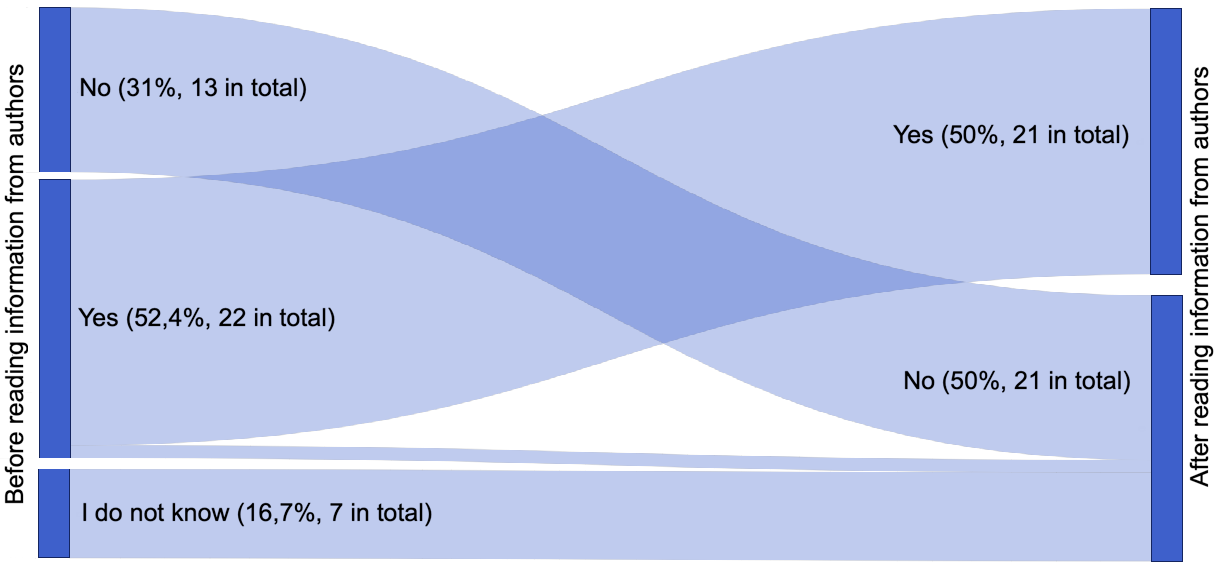}}
\caption{Concerns about data leak from project}
\label{fig:Q1617}
\end{figure}

As shown in Fig.~\ref{fig:Q1617} before reading the information more than half 22 (52,4\%) of the respondents have concerns about data leak from project, 13 (31\%) are not afraid and only 7 (16,7\%) do not know. After getting acquainted with the information the answers to 'Yes' and 'No' were the same amount 21 (50\%).

In the last open question respondents were pleased express their opinion. Analyzing all the answers, it was often stated that the tool can help and speed up the work when the code is not very complicated. It will be problematic to help implement more complex components. It can be very helpful for juniors or students in the learning process and gaining knowledge. Many interviewees had concerns about data leakage, copyright and security issues. Many of the respondents claimed that the tool may be helpful, but it will not take the work away, because the market requires specialists.

\section{Discussion and Future Work}

The main objective of this research was to study software developers’ experience and opinion about Github Copilot in the software development process. 2/3 of respondents have heard about the tool, although it is relatively new. Most of the developers used it for testing and out of curiosity. Although 85.7\% did not use the tool, after getting acquainted with it, its possibilities and potential, more than half would like to test it and believe that it seems to be helpful and speeding up their work. There are also those who are negative about such solutions and do not want to use them.
By analyzing the specialization chart far outweighs the answer 'It won't have an affect'. Backend is a more interesting case, because the 'Positive' and 'Negative' responses are on a similar level. On the other hand, for those cases where more input has a frontend (Frontend, Fullstack, Mobile), there is an upward trend towards the answer that it will have a positive impact. In the case of the division into the level of experience for Mid and Seniors 'It won't have an effect' it is several times higher than the other answers. There are only a few negative votes, but more positive votes. But the chart for Juniors stands out, because there are the most answers that it will affect negatively. Despite the authors' assurances, the developers' opinion on security with data transmission or data leak is ambiguous and most have concerns.

Many of the programmers have not used GitHub Copilot, which may be due to the long waiting time on the 'GitHub Copilot Waitlist' [5] and the lack of a work permission because of security and copyright. This seems like a good reason to survey a larger group of developers who used and tested the tool over a longer period of time. In this group, most of the respondents did it for several weeks.
For the most part, the tool was found to be interesting, worthy of attention and testing. It can be checked how the tool will affect on productivity by level of experience. This would be a division into two larger groups, for example Juniors, who would have the same tasks with and without GitHub Copilot. Security issues are often questionable and do not inspire confidence among developers. In order to check whether there is really no risk with sending data to the project e.g. an exploit, an interesting research would be to use it for a longer period of time on a larger group of people and check the results.

\section{Conclusion}

The main purpose of our research was to explore software developers' attitude to the new AI-pair programmer Github Copilot. Survey results show that opinions about this type of tools differs among developers. Most of them heard about Github Copilot before attending the survey but only 14,3\% used it. As far as usage of Github Copilot is concerned most applications referred to non commercial projects like testing the tool, portfolio, studies or open source projects.

The part concerning developers' attitude to Github Copilot indicates that there was low interest in using it. Only 38,1\% of participants were eager to test this tool though most of them had positive attitude to it and considered Github Copilot as helpful. As for the impact on future employment of programmers most answers pointed out a negative effect on juniors while other seniority levels are secure according to the majority.

Answers indicate that most of participants of the survey specialized in Backend programming. Fullstack and Frontend developers also made up a significant part of respondents. Only 21,4\% of participants were seniors, bigger groups were formed by Juniors and Mids, 42,9\% and 35,7\% respectively.

As for security issues the opinions of programmers are divided. A significant part of respondents has concerns about data sent into projects and data leak from projects. Even after reading the Github Copilot's authors assurance that data are sent safely not much participants changed their mind.

\vspace{12pt}


\begin{thebibliography}{00}
\bibitem{b1} Dominik Sobania, Martin Briesch and Franz Rothlauf, ``Choose Your Programming Copilot''
\bibitem{b2} Hammond Pearce, Baleegh Ahmad, Benjamin Tan, Brendan Dolan-Gavitt, Ramesh Karri, ``Asleep at the Keyboard? Assessing the Security of Github Copilot's Code Contributions''
\bibitem{b3} Neil A. Ernst and Gabriele Bavota, ``AI-driven Development Is Here: Should You Worry?''
\bibitem{b4} Raphael Jenni 'Machine learning for programming languages'.
\bibitem{b5} https://copilot.github.com/
\end{thebibliography}
\end{document}